\documentclass[12pt]{iopart}

\usepackage{iopams}
\usepackage{amssymb}
\usepackage{graphicx}
\usepackage{bm}
\usepackage{color} 

\def\la{{\langle}}
\def\ra{{\rangle}}

\newcommand{\beq}{\begin{equation}}
\newcommand{\eeq}{\end{equation}}
\newcommand{\beqa}{\begin{eqnarray}}
\newcommand{\eeqa}{\end{eqnarray}}

\begin{document}

\title{Qubit gates with simultaneous transport in double quantum dots}
\author{Yi-Chao Li$^{1,2}$, Xi Chen$^1$, J. G. Muga$^2$, E. Ya.  Sherman$^{2,3}$}
\address{$^{1}$ Department of Physics, Shanghai University, Shanghai 200444, Peoples Republic of China}
\address{$^{2}$ Department of Physical Chemistry, UPV/EHU, Apdo. 644, 48080 Bilbao, Spain}
\address{$^{3}$ IKERBASQUE Basque Foundation for Science, Bilbao, Spain}
\eads{
\mailto{jg.muga@ehu.eus},
\mailto{xchen@shu.edu.cn}}
\begin{abstract}
A single electron spin in a double quantum dot in a magnetic field is considered in terms of a four-level system.
By describing the  electron motion between the potential minima by spin-conserving tunneling and spin flip caused by a
spin-orbit coupling, we inversely engineer faster-than-adiabatic  state manipulation operations based on the geometry of
four-dimensional (4D) rotations. In particular, we show
how to transport a qubit among the quantum dots performing simultaneously a required spin
rotation.
\end{abstract}
\date{\today}

\maketitle

\section{INTRODUCTION}%
%
%
%
Device architecture based on electrons confined in coupled quantum dots \cite{Loss1998,Oosterkamp1998,Hu2000,Hu2001}
is considered as a potential and significant candidate for quantum computation and quantum information processing.
The advantages of this architecture are based on the facts that electron spin is a natural qubit with spin-up and
spin-down states, mature semiconductor technology may be used, and
long coherence times on the scale of microseconds have been achieved in these systems \cite{Bluhm2011,Veldhorst2014}.
Laboratories use electric, microwave or magnetic fields to manipulate spin states, performing $10^3\thicksim10^5$ operations in the
spin dephasing time \cite{Bluhm2011,Veldhorst2014,Nowack2007,McNeil2011,Baart2016_1,Baart2016_2}.

Scalability of quantum information devices is associated with several architectures having
the capability to transport qubits. In this paper
we  theoretically explore a four-level model for a spin in a double quantum dot (DQD) aiming at
the possibilities to implement fast qubit transport with simultaneous rotations.
We achieve this goal for arbitrary rotations by controlling the synchronized time dependence of interdot tunneling and spin-orbit coupling (SOC).
We inverse-engineer these time dependencies based on our recent work \cite{Yichao2018} on the control of four-level systems.
The method separates population control from control of the phases of the bare state basis \cite{Kang2017}.
Populations can be mapped onto a 4D sphere
so their evolution amounts to 4D transformations controlled by the rotation Hamiltonian
that may be engineered from the target state (in our case via isoclinic rotations and quaternions).
A full Hamiltonian can then be constructed from the rotation Hamiltonian to realize the desired phase changes.
Arbitrary state manipulations require full flexibility in the Hamiltonian, i.e., the possibility
to implement the different Hamiltonian matrix elements with specific time-dependences.
In the systems of interest, however,  there are constraints that hinder certain manipulations and transitions.
In particular, in this paper we examine the Hamiltonian structure
that corresponds to combined tunneling and SOC controllable couplings, and deduce the
possible transformations.

Spin-orbit coupling in semiconductors
consists of two main contributions due to the Dresselhaus- and the Bychkov-Rashba-effect.
The former is due to the bulk inversion asymmetry of material and the latter results from the structure
inversion asymmetry, produced, e.g., by the confining potential or an external electric field \cite{Fabian2007}. The practical
advantage of the Rashba coupling is the ability to manipulate it by an external electric field applied
across the semiconductor structure \cite{Nitta1997,Sawada2018}. The Rashba coupling
controlled by a high-frequency ac gate voltage \cite{AG2003}
provides  an effective method to control the spin states in short times \cite{Sadreev2013,Arrondo2013}.

This paper is organized as follows. In Section II, we introduce first the method
that parameterizes the time-dependent Hamiltonian and time evolution operator of a four-level system
by using isoclinic rotations and quaternions \cite{Yichao2018}. Then we
map the Hamiltonian of the spin in a DQD coupled by SOC and tunneling  onto this scheme.
In Section III, we apply the method developed in Section II to design the synchronized time dependences
of the control parameters to perform different qubit operations,
such as the interdot transport combined with spin rotations. Section IV provides discussion of the results and
their relation to other systems. Some details on the structure of the Hamiltonian are presented in the Appendix.
%
%

%
%
%
%
%
%
\section{ELECTRON IN A DOUBLE QUANTUM DOT: A 4D APPROACH}

\subsection{4D Hamiltonians and evolution operators}
The wave function of a four-level system
\beq
\psi(t)=\sum_{n=1}^{n=4}c_{n}(t)e^{i\varphi_{n}}|n\ra,
\eeq
where $c_n, \varphi_n$ are real amplitudes and phases (we set $\varphi_{1}=0$), and $\sum_{n}^4 c^2_n=1$,
can be decomposed as $|\psi(t)\ra=K(t)|\psi_r(t)\ra$, where
\beq
\psi_r(t)=\sum_{n=1}^{n=4}c_{n}(t)|n\ra
\eeq
is a vector on the surface of a 4D sphere, and the phase information is contained in
\beq
K(t)=\sum_{n=1}^{n=4}e^{i\varphi_{n}}|n\ra \la n|.
\eeq
The states $|\psi(t)\ra$ and $|\psi_r(t)\ra$ evolve via evolution operators $U(t)$ and $U_r(t)$ related by $U_r(t)=K^\dagger(t)U(t)K(0)$,
\beqa
&&|\psi(t)\ra=U(t)|\psi(0)\ra,
\nonumber \\
&&|\psi_r(t)\ra=U_r(t)|\psi_r(0)\ra,
\eeqa
where we set the initial time as $0$. Accordingly, the rotation-related Hamiltonian in the Hilbert space is defined as
\beq
\label{Hr}
H_r(t)=i \hbar \dot{U}_r(t) U_r^\dagger (t),
\eeq
and the total Hamiltonian  is
\beqa
\label{H}
H(t)&=&i \hbar \dot{U}(t) U^\dagger (t)
\nonumber\\
&=&K(t)H_r(t)K^\dagger(t)+i \hbar \dot{K}(t)K^\dagger (t).
\eeqa
To engineer $H_r$ for a specific rotation, it is convenient to express first a
general 4D rotation matrix as a  product of two isoclinic
rotation matrices \cite{Thomas2014,Thomas2017}:
\beq
\label{rotation}
U_r(t)=
\left[
 \begin{array}{cccc}
q_w&-q_x&-q_y&-q_z\\
q_x&q_w&-q_z&q_y\\
q_y&q_z&q_w&-q_x\\
q_z&-q_y&q_x&q_w\\
\end{array}
\right]
\left[
 \begin{array}{cccc}
p_w&-p_x&-p_y&-p_z\\
p_x&p_w&p_z&-p_y\\
p_y&-p_z&p_w&p_x\\
p_z&p_y&-p_x&p_w\\
\end{array}
\right],
\eeq
where $q_i$ and $p_j$ are components of two unit quaternions
$q=q_w+q_x \textbf{\textit{i}}+ q_y \textbf{\textit{j}}+ q_z \textbf{\textit{k}}$ and
$p=p_w+ p_x \textbf{\textit{i}}+ p_y \textbf{\textit{j}}+ p_z\textbf{\textit{k}}$. We shall parameterize
 them in terms of generalized 4D spherical angles \cite{Sommerfeld1949,Muga1995},
\beqa
\label{angles}
q_w(t)&=&\cos\gamma_1(t),
\nonumber\\
q_x(t)&=&\sin\gamma_1(t) \cos\theta_1(t),
\nonumber\\
q_y(t)&=&\sin\gamma_1(t) \sin\theta_1(t) \cos\phi_1(t),
\nonumber\\
q_z(t)&=&\sin\gamma_1(t) \sin\theta_1(t) \sin\phi_1(t),
\nonumber\\
p_w(t)&=&\cos\gamma_2(t),
\nonumber\\
p_x(t)&=&\sin\gamma_2(t) \cos\theta_2(t),
\nonumber\\
p_y(t)&=&\sin\gamma_2(t) \sin\theta_2(t) \cos\phi_2(t),
\nonumber\\
p_z(t)&=&\sin\gamma_2(t) \sin\theta_2(t) \sin\phi_2(t),
\eeqa
where $0\le \phi_{1,2}\le 2\pi$, $0\le \theta_{1,2},\gamma_{1,2}\le \pi$.
Thus by using $U(t)=K(t)U_r(t)K^{\dagger}(0)$ and $ H(t)=K(t) H_r(t) K^{\dag}(t) + i \dot{K}(t) K^{\dag}(t) $,
we find the parameterized forms for the evolution operator $U(t)$ and the Hamiltonian $H(t)$.
The  explicit expressions are lengthy, and will not be reported here.
\subsection{Single electron in a double quantum dot}
%
%
\begin{figure}[t]
\begin{center}
\scalebox{0.5}[0.5]{\includegraphics{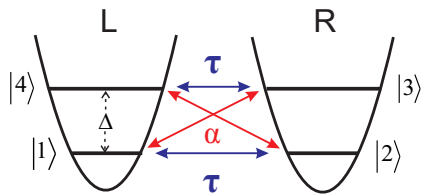}}
\caption{ Schematic diagram of single electron states in a double quantum dot.}
\label{fig1}
\end{center}
\end{figure}
%
Consider a single electron spin in a semiconductor DQD, for example made of silicon or GaAs,
with tunneling and Rashba spin-orbit coupling, as shown in Fig. \ref{fig1}.
We use a bare basis of spin up and down states localized in
each well, numbered
as $|\psi_{L\downarrow}\ra=|1\ra$, $|\psi_{R\downarrow}\ra=|2\ra$, $|\psi_{R\uparrow}\ra=|3\ra$, $|\psi_{L\uparrow}\ra=|4\ra$.
Following the derivation in \ref{app1}, and after
an diagonal energy shift of $-{\Delta}/2,$ the Hamiltonian of this system (see (\ref{totalH})) can be written as
\beq
\label{H0}
H_0(t)=\hbar
\left[  \begin{array}{cccc}
0 & \tau (t) & \alpha (t) & 0\\
\tau (t) & 0 & 0 & -\alpha (t)\\
\alpha^{*}(t) & 0 & \Delta & \tau (t) \\
0 & -\alpha^{*}(t) & \tau (t) & \Delta \\
\end{array} \right].
\eeq
Here $\tau(t)$ represents the tunneling coupling between the two quantum dots,  $\alpha(t)$ is the Rashba
coupling, and
$\Delta$ is a Zeeman splitting. All these quantities have dimensions of frequency.
Following the approach of Mal'shukov \textit{et al.} \cite{AG2003}, we consider the time-dependent Rashba coupling in the complex
form $\alpha(t)=\alpha_0+\alpha_1(t) e^{i \omega t}$.

The Hamiltonian structure corresponds topologically to a diamond-configuration \cite{Yichao2018},
which, in the parametric expression of $H(t)$ we may impose with the conditions
\beq
\label{diamond}
\dot{\theta}_1=\dot{\theta}_2=\dot{\phi}_1=\dot{\phi}_2=0,\ \phi_1=\phi_2=0,
\eeq
see Fig. \ref{fig2}.
Specifically, after substituting (\ref{diamond}) in the parameterized form of ({\ref{H}), the Hamiltonian $H(t)$ acquires the corresponding form
%
\begin{figure}[t]
\center
\scalebox{0.5}[0.5]{\includegraphics{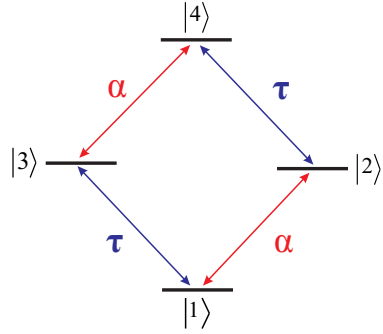}}
\caption{Schematic links between the bare states in Fig. 1 showing the ``diamond'' structure of the transition.
Here the positions of the levels are not related to their energies.}
\label{fig2}
\end{figure}
%
%
\beqa
H(t)&=&
\label{H2}
\hbar \{-\dot{\varphi}_2(t)|2\ra \la 2|-\dot{\varphi}_3(t)|3\ra \la 3|-\dot{\varphi}_4(t)|4\ra \la 4|
\nonumber\\
&-& i [e^{-i \varphi_2(t)}(\dot\gamma_1(t)\cos{\theta_1}+\dot\gamma_2(t)\cos{\theta_2})]|1\ra \la 2|
\nonumber\\
&+&e^{-i \varphi_3(t)}(\dot\gamma_1(t)\sin{\theta_1}+\dot\gamma_2(t)\sin{\theta_2})|1\ra \la 3|
\nonumber\\
&+&e^{i [\varphi_2(t)-\varphi_4(t)]}(-\dot\gamma_1(t)\sin{\theta_1}+\dot\gamma_2(t)\sin{\theta_2})|2\ra \la 4|
\nonumber\\
&+&e^{i [\varphi_3(t)-\varphi_4(t)]}(\dot\gamma_1(t)\cos{\theta_1}-\dot\gamma_2(t)\cos{\theta_2})|3\ra \la 4|]\}
\nonumber\\
&+&{\rm H.c.}
\eeqa
To make $H_0(t)$ and $H(t)$ fully consistent, we further fix the angles as
%
%
\beqa
\label{condition}
&&\dot{\theta}_2=\theta_2=\dot{\gamma}_2=\gamma_2=0, \nonumber\\
&&\varphi_2(t)=-\frac{\pi}{2}, \varphi_3(t)=-\Delta t+\frac{\pi}{2}, \varphi_4(t)=-\Delta t.
\eeqa
Then (\ref{H2}) gives
\beq
\eqalign{
H(t)&=\hbar\{
\dot{\gamma}(t)[\cos\theta(|1\ra \la 2|+|3\ra \la 4|) -e^{i \Delta t} \sin\theta (|1\ra \la 3|-|2\ra \la 4|)]
\\
& +\Delta(|3\ra \la 3|+|4\ra \la 4|)\}+{\rm H.c.},
}
\eeq
where we have simplified the notation as $\gamma(t)=\gamma_1(t)$, $\theta=\theta_1$.
Now we may impose $H_0(t)=H(t)$, as they have the same structure, to find the following relations between
control functions and auxiliary angles,
\beqa
\label{tandr}
&&\tau(t)=\dot{\gamma}(t) \cos\theta,\nonumber\\
&&\alpha(t)=-e^{i \Delta t}\dot{\gamma}(t)\sin\theta,
\eeqa
which implies $\alpha_0=0$, $\alpha_1(t)=-\dot{\gamma}_1(t)$, $\omega=\Delta$
(i.e., the external bias is in resonance with the Zeeman frequency),
and $\theta$ can be considered as a coupling mixing angle.
Under the conditions stated in Eqs. (\ref{condition}), the parameterized time-evolution operator becomes
\begin{small}
\beq \label{u(t)}
\fl
\eqalign{U(t)= \\
\left[  \begin{array}{cccc}
 \cos \gamma(t) & -i \cos \theta \sin \gamma(t) & i \sin \gamma(t) \sin \theta & 0 \\
 -i \cos \theta \sin \gamma(t) & \cos \gamma(t) & 0 & -i \sin \gamma(t) \sin \theta \\
 i e^{-i \Delta t } \sin \gamma(t) \sin \theta & 0 & e^{-i \Delta t } \cos \gamma(t) & -i e^{-i \Delta t } \cos \theta \sin \gamma(t) \\
 0 & -i e^{-i \Delta t } \sin \gamma(t) \sin \theta & -i e^{-i \Delta t } \cos \theta \sin \gamma(t) & e^{-i \Delta t } \cos \gamma(t)
\end{array} \right]}.
\eeq
\end{small}
%
We impose the boundary condition $\gamma(0)=2n\pi,\, n=\ldots,-2,-1,0,1,2,\ldots$, to guarantee $U(0)=1$ at the initial time.
\section{Applications}
\label{sec2}
\subsection{Qubit preparation}
\label{subsec1}
Assume that the four-level system is initialized in state $|1\ra$ on the left well and the
objective is to prepare from it an arbitrary qubit in the right well
encoded in levels $|2\ra$ and $|3\ra$.
Besides the conditions in (\ref{condition}), we set $\psi(0)=(1,0,0,0)$ and $\psi(T)=(b_1,b_2,b_3,b_4)$,
where $T$ is the duration time and $b_n$ are final complex amplitudes which satisfy $\sum_{n}|b_{n}|^2=1$.
By using $\psi(T)=U(T)\psi(0)$, we have
\begin{eqnarray}
\label{bn}
b_1&=& \cos \gamma(T),
\nonumber\\
b_2&=&-i \cos \theta \sin \gamma(T),
\nonumber\\
b_3&=& i e^{-i \Delta T } \sin \gamma(T) \sin \theta,
\nonumber\\
b_4&=&0.
\end{eqnarray}
We can transfer $|1\ra$ to any bare state except $|4\ra$, or to arbitrary superpositions of $|2\ra$ and $|3\ra$
(i.e., any qubit on the right well) by imposing $\gamma(T)=(2n+1)\pi/2$, $n=0,\pm1,...$.

As an example we shall perform a state transfer to
$b_1=0,b_2=1/2,b_3=e^{i{\pi}/{2}}\sqrt{3}/2,b_4=0$.
Equation (\ref{bn}) with
\beqa
&&\theta=-\frac{\pi}{3},\;\;\,\,\,\, T=\frac{3\pi}{2\Delta},\nonumber\\
&&\gamma(0)=0,\;\;\,\, \gamma(T)=\frac{\pi}{2},\nonumber\\
&&\dot{\gamma}(0)=0,\;\;\,\, \dot{\gamma}(T)=0,
\eeqa
corresponds to the desired final state $|\psi(T)\ra=-i(|2\ra+e^{i\pi/2}\sqrt{3}|3\ra)/2$ within
an irrelevant global phase factor. An Ansatz for $\gamma(t)$ consistent with the above boundary conditions is
\beq
\label{ansatz}
\gamma(t)=\frac{\pi}{4}\left[1-\cos \left(\frac{\pi t}{T}\right)\right].
\eeq
The resulting tunneling and Rashba SOC are calculated from (\ref{tandr}) as
\begin{eqnarray}
&&\tau(t)=\frac{\pi^2}{4T}\sin\left(\frac{\pi t}{T}\right) \cos\theta, \nonumber \\
&&\alpha(t)=-e^{i \Delta t}\frac{\pi^2}{4T}\sin\left(\frac{\pi t}{T}\right) \sin\theta,
\end{eqnarray}
with the characteristic values $\tau,\alpha \propto {1}/{T}$. We can prepare a qubit with an arbitrary relative phase
by adjusting $\Delta$ and the operation time $T$ as long as the tunneling and SOC are experimentally feasible.
We plot the time-dependence of the tunneling matrix elements, Rashba SOC, and populations evolution of all bare states
in Fig. \ref{fig3} with parameters corresponding to the $g^{*}-$factor of electron
in GaAs ($g^{*}=-0.44$) with $B=100$ mT, $\Delta \approx 2\pi \times 0.5$ GHz,  and $T=3\pi/(2\Delta)=1.5$ ns  \cite{Petta2005}.

\begin{figure}[t]
\center
\scalebox{0.7}[0.7]{\includegraphics{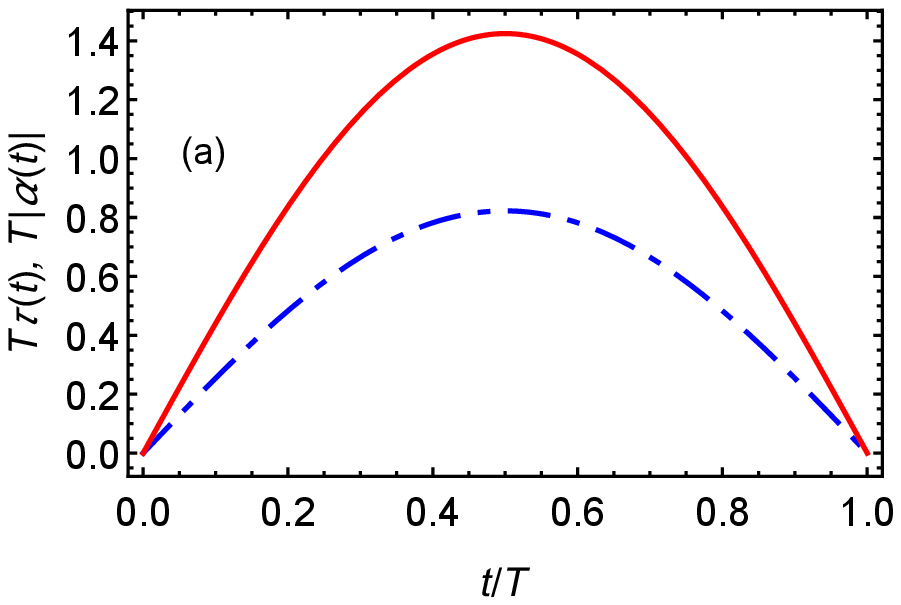}}
\scalebox{0.7}[0.7]{\includegraphics{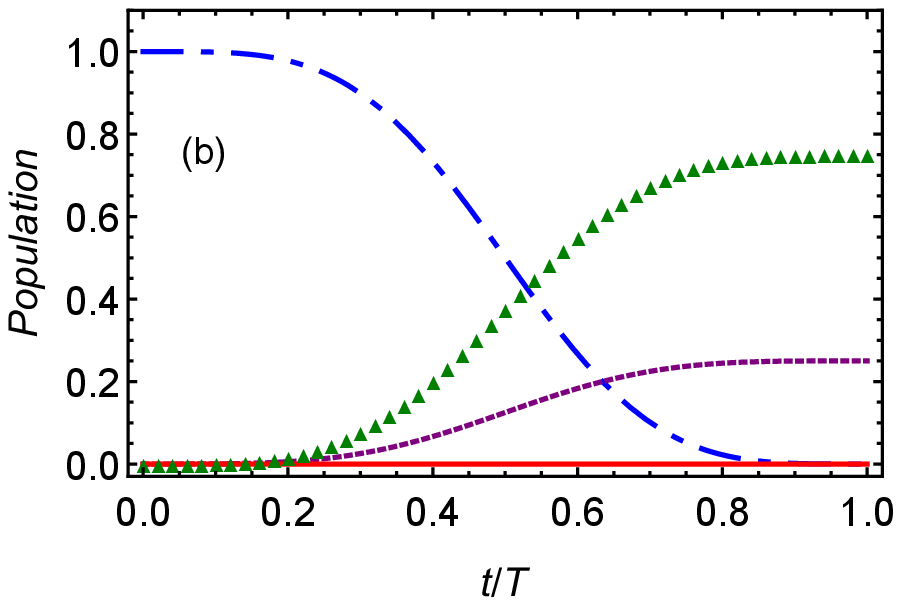}}
\caption{ (a) $T\tau(t)$ (dot-dashed blue line) and $T|\alpha(t)|$ (solid red line).
 (b) Populations of $|1\ra$ (dot-dashed blue line),
$|2\ra$ (dashed purple line), $|3\ra$ (green triangles) and $|4\ra$ (solid red line)
for the transfer from  $|1\ra$ to $-i(|2\ra+e^{i\pi/2}\sqrt{3}\ |3\ra)/2$.
Parameters:  $B=100$ mT, $\Delta \approx 2\pi \times 0.5$ GHz, and $T=1.5$ ns.}
\label{fig3}
\end{figure}
%
%
\subsection{Qubit transport and rotation}
\label{subsec2}
Our method may be applied to transport the qubit from one dot to the other applying
simultaneously some qubit rotation, i.e., to produce an arbitrary gate.
Suppose we have already prepared a qubit in the left dot in an arbitrary
superposition of $|1\ra$ and $|4\ra$ as $\psi(0)=\cos \chi |1\ra+ e^{i \mu}\sin \chi |4\ra$, where $\chi$
is the initial amplitude mixing angle and $\mu$ is the initial relative phase. The corresponding
general final state with the unitary evolution operator (\ref{u(t)}) is
given by the amplitudes
\begin{eqnarray}
\label{bn2}
b_1&=& \cos\chi \cos\gamma(T),
\nonumber\\
b_2&=& A e^{i \zeta_A},
\nonumber\\
b_3&=& B e^{i (\zeta_B-\Delta T)},
\nonumber\\
b_4&=& \sin\chi \cos\gamma(T)e^{i (\mu-\Delta T)},\\\nonumber
\end{eqnarray}
where
%
\beqa
A&=&\frac{\sin\gamma(T)}{\sqrt{2}}\sqrt{1+\cos2\chi\cos 2\theta + \cos\mu \sin2\chi \sin2\theta},\nonumber\\
B&=&\frac{\sin\gamma(T)}{\sqrt{2}}\sqrt{1-\cos2\chi\cos 2\theta-\cos\mu \sin2\chi \sin2\theta},\nonumber\\
\zeta_A&=&-\arctan\left(\cot\mu + \cot\theta  \frac{\cot\chi}{\sin\mu}\right),\nonumber\\
\zeta_B&=&-\arctan\left(\cot\mu - \tan\theta  \frac{\cot\chi}{\sin\mu}\right).
\eeqa
%
We can inversely calculate the coupling mixing angle $\theta$ under the condition that $\gamma(T)= n\pi/2, n=1,3,5,\cdots$,
so that the amplitudes $b_1$ and $b_4$ vanish, and for given desired final real amplitudes $A$ and $B$ we obtain:
\beq \label{theta}
\fl
\theta=
\pm \arccos
\sqrt{
\frac{A^2 \cos2\chi +\sin^2\chi(1-2\cos^{2}\chi\sin^{2}\mu) + 2 \, S \,\cos\mu \sin2\chi }
{1-\sin^{2}2\chi\sin^2\mu}
},
\eeq
where
\beq
S=\sqrt{ 4 A^2 B^2 -\sin^2 \mu \sin^2 2\chi}.
\eeq
Notice that there is still a degree of freedom to control the final relative phase of the qubit on the right.
Suppose our target relative phase  is $\lambda=\zeta_B-\zeta_A-\Delta T$.
By adjusting the operation time as  $T=(\zeta-\lambda)/\Delta$,
where $\zeta=\zeta_B-\zeta_A$, the process produces the desired relative phase $\lambda$.
Now we can consider two examples of application of (\ref{theta}).

\begin{figure}[t]
\center
\scalebox{0.7}[0.7]{\includegraphics{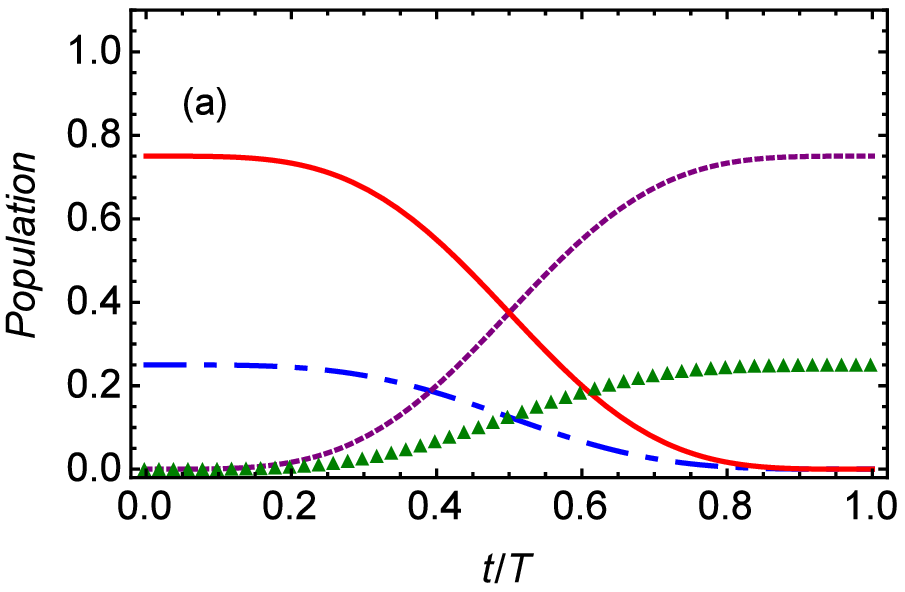}}
\scalebox{0.7}[0.7]{\includegraphics{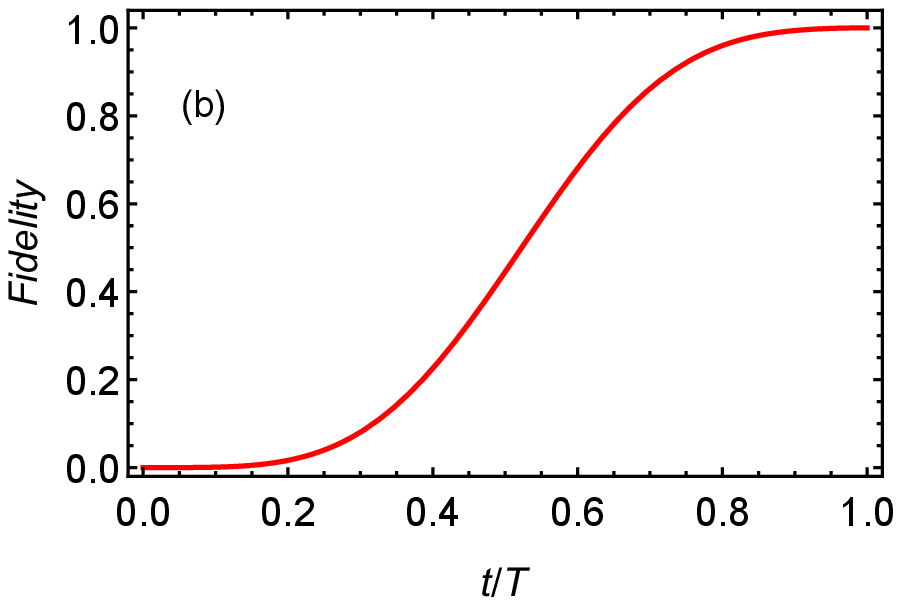}}
\caption{(a) Transient bare populations for the transition from $(|1\ra+e^{i \pi/4}\sqrt{3}\ |4\ra)/2$ to $(e^{i \pi/4}\sqrt{3}\ |2\ra+|3\ra)/2$.
(b) Fidelity versus time $t$. The parameters are $\gamma(T)=\pi/2$, $\mu=\pi/4$, $\chi=\pi/3$, $T=0.5$\ ns,
$\Delta=2\pi \times 0.5$ GHz
and $\theta=0.685$ (as calculated by (\ref{theta})), lines are $|1\ra$ (dot-dashed blue line),
$|2\ra$ (dashed purple line), $|3\ra$ (green triangles) and $|4\ra$ (solid red line).
}
\label{fig4}
\end{figure}
\
\begin{figure}[t]
\center
\scalebox{0.85}[0.85]{\includegraphics{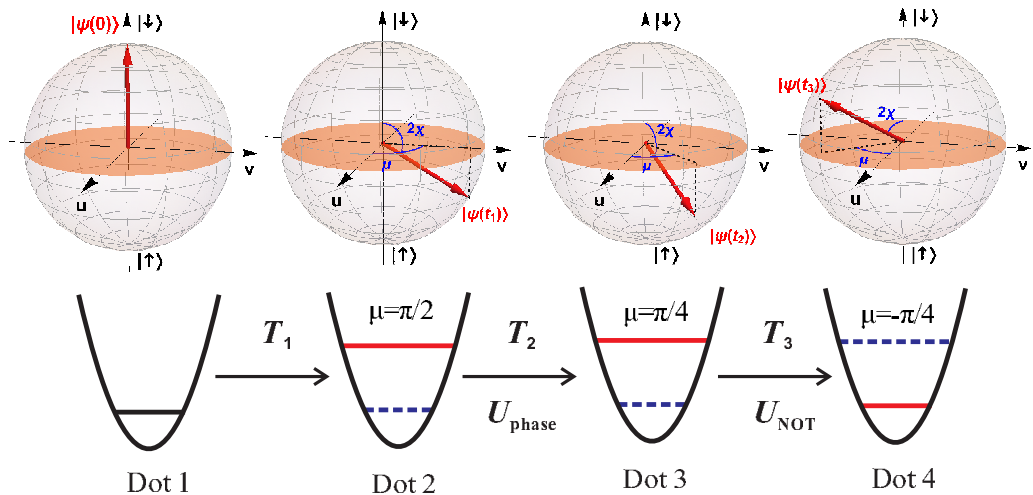}}
\caption{Scheme for qubit operations in a chain of quantum dots. The system is initialized in state $|\downarrow\ra$ in
Dot 1 at time $t=0$; then, from this initial state, a qubit is prepared in Dot 2 at a time $t_1$
(the duration of the process is $T_1=t_1$). This qubit is transported to Dot 3 with an additional
relative phase $\pi/4$ (process duration $T_2=t_2-t_1$). Finally, a ``NOT gate with transport''
operation is applied to flip and transport the qubit to Dot 4 in a process with duration $T_3=t_3-t_2$.
The upper figures represent each qubit on the corresponding Bloch spheres.}
\label{fig5}
\end{figure}

%
%


\textit{Example 1: transport and phase gate.} We assume that $A=\cos\chi$ and $B=\sin\chi$ and
substitute them into (\ref{theta}) to get $\theta=0$, which
means that $\tau=\dot{\gamma}(t)$ and $\alpha=0$. The final state is calculated as
\begin{eqnarray}
\label{bn_phase}
b_1&=& \cos\chi \cos \gamma(T),
\nonumber\\
b_2&=& -i  \sin \gamma(T)\cos\chi,
\nonumber\\
b_3&=& -i e^{i (\mu-\Delta T) } \sin \gamma(T) \sin\chi,
\nonumber\\
b_4&=& \sin \chi \cos \gamma(T)e^{-i\Delta T}.
\end{eqnarray}
%
By letting $\gamma(t)$ evolve from 0 to $\pi/2$, the qubit is transported from left
to right and rotated by a relative phase factor $e^{-i \Delta T }$.

\textit{Example 2: transport and NOT gate.} The NOT gate with transport swaps the
amplitudes between up and down states, so we set $A=\sin\chi$ and $B=\cos\chi$ in (\ref{theta}) to get
\beq
\label{theta2}
\theta=\frac{\cos\mu \sin2\chi}{\sqrt{1-\sin^{2}2\chi\sin^2\mu}}.
\eeq
We also impose $\lambda=-\mu$, so we fix the operation time as $T=(\zeta+\mu)/\Delta$.

The population transfer of the NOT gate+transport  operation is depicted in Fig. \ref{fig4}.
We use the same Ansatz for $\gamma(t)$ in (\ref{ansatz}) to keep $\gamma(T)=\pi/2$ and
choose for the example $\mu=\pi/4$, $\chi=\pi/3$.
The population is successfully swapped in Fig. \ref{fig4} (a), and Fig. \ref{fig4} (b)
shows the fidelity  $F=|\la \psi_{\rm trg}|\psi(t)\ra|^2$, where the target state is
$|\psi_{\rm trg}\ra=\sin\chi e^{i\mu}|2\ra+\cos\chi|3\ra$.

\section{Discussion}%

By applying an approach based on four-dimensional rotations, we studied electron charge and spin motion in tunneling-
and spin-orbit coupled quantum dots. By a proper synchronization of their time-dependences,
we inversely engineered the tunneling and spin-orbit coupling matrix
elements to achieve spin transport with simultaneous single qubit rotations in quantum information
transformations such as the qubit preparation and $U_{\rm phase}$ and $U_{\rm NOT}$ gates. %
In a chain of quantum dots, these transport+rotation operations may be applied sequentially for a long-distance qubit
transfer in a multi-dot architecture, where the ability of a coherent spin transfer has
been recently demonstrated \cite{Flentje2017,Fujita2017}.
Figure \ref{fig3} demonstrates these processes for a particular sequence starting in Dot 1 and ending in Dot 4.
We point out that this technique can also be applied to heavy-hole systems, where  the
control of the hole spin via the tunneling and strong SOC has been demonstrated for silicon-based double quantum dots \cite{Bogan2017}.
In addition, a similar approach can be used to design the spin and mass transport of cold atoms
in optically produced potentials \cite{Kartashov2018}.

\ack
{We acknowledge this paper is supported by the China Scholarship Council(CSC), by the National Natural Science Foundation of China (Grant No. 11474193), the Shuguang program (Grant No. 14SG35), the program of Shanghai Municipal Science and Technology Commission (Grants No. 18010500400 and No. 18ZR1415500), by the Spanish Ministry of Economy, Industry and Competitiveness (MINECO) and the European Regional Development Fund FEDER through Grant No. FIS2015-67161-P (MINECO/FEDER, UE), and the Basque Government through Grant No. IT986-16.}

\appendix
\section{Four-level Hamiltonian of a double quantum dot}
\label{app1}
We consider a single electron in a double quantum dot modeled by a one-dimensional Hamiltonian,
as can be realized in nanowire-based systems \cite{Nadj2010}, where the electron is tightly confined
in the perpendicular directions, as
\beq
H_{\rm DQD}(t)=\frac{\widehat{p}^2}{2m}+V(x).
\eeq
Here the first term is the kinetic energy with $\widehat{p}=-i\hbar \partial/\partial x$  and $m$ is the electron effective mass
(e.g.,  in GaAs $m \approx 0.067$ of the free electron mass). We assume that $V(x)$ is a spatially symmetric potential ($V(x)=V(-x)$)
with two equivalent minima at points $x_{0}>0$ and $-x_{0}$ and choose the basis for the tunneling-related
Hamiltonian as two approximate states of this potential, localized in the vicinity of the
points $-x_{0}$ and $x_{0},$ which we will denote as $\widetilde{\psi}_{L(R)}(x)$ \cite{Li2017,Burkard1999}, respectively.
The Hamiltonian in the basis becomes
\beq
H_{\rm tun}=\hbar
\left[  \begin{array}{cccc}
0 & \tau\\
\tau & 0
\end{array} \right],
\eeq
where $\tau$ is the tunneling rate between the two quantum dots determined by deviation of the total potential
$V(x)$ from its shape in the vicinity of the minima \cite{Ashcroft}.
Thus, by modifying $V(x)$ by a time-dependent external field, one can produce time-dependent $\tau(t).$

A magnetic field $\mathbf{B}=(0,0,-B)$ along the $z-$axis
causes the Zeeman spin splitting corresponding to the Hamiltonian
$H_{Z}=-g^{*}\mu_{B}\sigma_{z}B/2$, where $g^{*}$ is the
conduction band Land\'{e} factor, $\mu_B$ is the Bohr magneton, and the level
splitting is $\Delta=\mu_{B}|g^{*}B|$ \cite{Bfield}.
In the basis of $\sigma_{z}-$representation, the eigenstates of $H_{Z}$ are
given by: $|\psi_\uparrow\ra=(1,0)^{\rm T}$ and $|\psi_\downarrow\ra=(0,1)^{\rm T}$.

The one-dimensional Rashba spin-orbit coupling is represented as
\beq
H_{\rm SOC}=\frac{\widetilde{\alpha}}{\hbar}\widehat{p}\sigma_{y},
\eeq
where the $\widetilde{\alpha}$ is the corresponding coupling parameter.

We define the full four-state basis of a single electron in the DQD as
\beq
|L(R)_{\uparrow(\downarrow)} \ra =|\psi_{L(R)}\ra \otimes |\psi_{\uparrow(\downarrow)}\ra
\eeq
and obtain nonzero coupling Rashba terms calculated as (with $\widehat{k}\equiv\widehat{p}/\hbar$)
\beqa
\hspace{-0.5cm}\la L_{\downarrow}|H_{R}|R_{\uparrow} \ra&=& \la \downarrow | \sigma_y |\uparrow \ra \la L| \widehat{k} |R\ra=-i \widetilde{\alpha} \la L| \widehat{k} |R\ra,
\nonumber \\
\hspace{-0.5cm}\la R_{\downarrow}|H_{R}|L_{\uparrow} \ra&=& \la \downarrow | \sigma_y |\uparrow \ra \la R| \widehat{k} |L\ra=-i \widetilde{\alpha} \la R| \widehat{k} |L\ra,
\eeqa
and the Hermitian conjugate terms. The diagonal
elements are all zero because  $\la L|\widehat{k}| L\ra=\la R|\widehat{k}| R\ra=0$ and $\la L|\widehat{k}| R\ra=-\la R|\widehat{k}| L\ra$.
We finally find in the basis $\{|L_{\downarrow}\ra, |R_{\downarrow}\ra, |R_{\uparrow}\ra , |L_{\uparrow}\ra\},$
corresponding to the  $\{|1\ra, |2\ra, |3\ra , |4\ra\}$ basis of the main text
\beq
H_R=\hbar
\left[  \begin{array}{cccc}
0&0&\alpha&0\\
0&0&0&-\alpha\\
\alpha^{*}&0&0&0\\
0&-\alpha^{*}&0&0\\
\end{array} \right],
\eeq
where  $\alpha\equiv -i \widetilde{\alpha} \la L| \widehat{k} |R\ra$.

For symmetric $V(x)$, the full Hamiltonian in the $\{|1\ra, |2\ra, |3\ra , |4\ra\}$ basis of the main text,
taking into account spin-conserving and spin-flip tunnelings acquires the form
\beq
\label{totalH}
H_0=\hbar
\left[  \begin{array}{cccc}
-{\Delta}/{2} & \tau(t) & \alpha (t)& 0
\\
\tau(t) &-{\Delta}/{2} & 0 & -\alpha(t)
\\
\alpha^{*}(t) & 0 & {\Delta}/{2} & \tau(t)
\\
0 & -\alpha^{*}(t)  & \tau(t) & {\Delta}/{2}
\\
\end{array} \right].
\eeq
The time dependence $\alpha(t)$ in (\ref{totalH}) comes from two main sources: time-dependent $\widetilde{\alpha}$ due to ac external bias
and time-dependent overlap of the wave functions localized near the left ($-x_{0}$) and right ($x_{0}$) minimum of the potential
$V(x).$
%


\begin{thebibliography}{99}
\bibitem{Bfield} Taking into account that we consider system with $g^{*}<0$, we take $\mathbf{B}$ antiparallel to the $z-$ axis
for consistency with the main text.
\end{thebibliography}
\end{document}